\documentclass[a4paper,10pt]{article}    
\usepackage{latexsym, amssymb}
\newcommand{\be}{\begin{equation}}
\newcommand{\ee}{\end{equation}}
\newcommand{\bea}{\begin{eqnarray}}
\newcommand{\eea}{\end{eqnarray}}

\catcode`@=11
\def\@sim#1#2{\setbox0=\hbox{$\sim$}\lower.9\ht0\vbox{\baselineskip0pt
          \lineskip0.1ex\ialign{$\m@th#1\hfill##\hfill$\crcr#2\crcr
          \sim\crcr}}}
\def\lsim{\mathrel{\mathpalette\@sim<}}
\def\gsim{\mathrel{\mathpalette\@sim>}}
\catcode`@=12

\newcommand{\artref}[5]{{\rm #1}, {\rm #2} {\bf #3}, {\rm #4 (#5)}}
\newcommand{\preprintref}[4]{{\rm #1}, {\rm #2}, {\rm #3}, {\rm #4}}

\newcommand{\bookref}[3]{{\rm #1}, {\it #2} {\rm (#3)}}

\def\B{B}
\def\Z{{\cal Z}}
\def\W{{\cal W}}
\def\gp{{g^\prime}}
\def\gs{g^2}
\def\gps{\gp^2}
\def\p{\partial}

\def\lag{{\cal L}}

\def\PT{{\rm PT}}

\def\tU{\widetilde U}

\def\order{{\cal O}}

\def\G{{\cal G}}  % Propagators

\def\ers{{e_{\mathrm{res}}^2}}

\def\d{{\rm d}}

\def\phase{{\varphi}}

\def\op{L}

\def\low{l}

\def\ourop{{\cal O}}  
\def\ourlow{l}

\def\MHps{M_{H,p}^2}  
\def\MWps{M_{W,p}^2}

\def\bU{{U}}
\def\bB{{B}}
\def\bW{{\W}} 
\def\bZ{{\Z}}

\def\bJ{{J}}
\def\bj{{j}}
\def\bK{{K}}
\def\bd{{d}}

\def\trho{{\bar \rho}}

\def\pole{\Lambda_\varepsilon}

\def\bv{{v}}
\def\bg{{g}}

\def\bgs{{{g}^2}}
\def\bgps{{{\bg^\prime}{}^2}}
\def\bc{{c}}
\def\bs{{s}}
\def\bes{{{e}^2}}

\def\Aslash{A\hbox to 0mm{\kern-2.3mm/\hss}}
\def\Dslash{D\hbox to 0mm{\kern-5mm/\hss}} 
                                           
\def\dslash{\partial\hbox to 0mm{\kern-2.2mm/\hss}}
\def\qslash{q\hbox to 0mm{\kern-1.75mm/\hss}}

\def\bu{{u}}
\def\bdf{{d}}

\begin{document}

\def\thefootnote{\fnsymbol{footnote}} 

\vspace*{-3.6cm}  
\begin{flushright}
LC-TH-1999-018 
\\
CPT-99/P.3910
\\
DESY 99-200
\\
hep-ph/9912472
\\
March 2000 
\end{flushright}

\vspace*{0.1cm}         
\begin{center}
{\Large {\bf The electroweak chiral Lagrangian
revisited\footnote{Based on talks presented at the 2nd ECFA/DESY Study
on Physics and Detectors for a Linear Electron-Positron Collider,
Oxford, United Kingdom, 20-23 March 1999 and at the 33rd International
Symposium Ahrenshoop on the Theory of Elementary Particles, Buckow,
Germany, 23-28 August 1999. To appear in the Proceedings of the 2nd
ECFA/DESY Study on ``Physics Studies for a Future Linear Collider'',
DESY report 123F (R.\ Heuer, F.\ Richard, P.\ Zerwas, eds.).}
}}\\[0.4cm]
    Andreas Nyf\/feler\\[0.2cm]
    DESY, Platanenallee 6, D-15738 Zeuthen, Germany\\%[0.1cm]
    and\\%[0.1cm]
    Centre de Physique Th\'{e}orique, CNRS-Luminy, Case 907, 
    F-13288 Marseille Cedex 9, France\footnote{Present address;
    e-mail: nyf\/feler@cpt.univ-mrs.fr }\\[0.5cm]

\begin{abstract}
Using a manifestly gauge-invariant approach we show that the set of
low-energy constants in the electroweak chiral Lagrangian currently
used in the literature is redundant. In particular, by employing the
equations of motion for the gauge fields, one can choose to remove two
low-energy constants which contribute to the self-energies of the
gauge bosons. The relation of this result to the experimentally
determined values for the oblique parameters $S,T$ and $U$ is
discussed. We then evaluate the matching relation between
gauge-invariant Green's functions in the full and the effective theory
for the case of the Standard Model with a heavy Higgs boson and
compare the results for the independent low-energy constants with
those for a simple Technicolor model. Since the pattern of the
low-energy constants is very different in these two models it may be
misleading to mimic any strongly interacting symmetry breaking sector
by a heavy Higgs boson. From our investigation we conclude that
current electroweak precision data do not really rule out such
strongly interacting models.
\end{abstract}
\end{center}

%---------------------------------------------------------------------------

\renewcommand{\thefootnote}{\arabic{footnote}}
\setcounter{footnote}{0}

\vspace*{-3mm}
\section{Introduction}

In Ref.~\cite{SM_gaugeinv} we introduced a manifestly gauge-invariant
approach to the bosonic sector of the Standard Model which deals with
gauge-invariant Green's functions and provides a method to evaluate
the corresponding generating functional without fixing a gauge. In
Ref.~\cite{EWChPT_gaugeinv} we employed this new approach to reanalyze
the electroweak chiral Lagrangian~\cite{EWChPT_early, EWChPT} in order
to investigate two issues related to gauge invariance.  We present
here the main results of our investigation and refer to
Refs.~\cite{SM_gaugeinv, EWChPT_gaugeinv} for a discussion of the
problems encountered with gauge invariance when studying these issues
in the usual approach to gauge theories, for the presentation of our
solution to these problems and for all the details of the derivation.
A more complete overview of the literature can be found in
Refs.~\cite{SM_gaugeinv, EWChPT_gaugeinv} as well.

The first issue concerns the determination of the number of
independent low-energy constants in the effective Lagrangian by
employing the equations of motion to eliminate redundant terms. This
is important for any general analysis of the data that will be
measured at future colliders like the LHC or a linear collider, if the
underlying dynamics of electroweak symmetry breaking is strongly
interacting. See Ref.~\cite{Han} for a recent overview.  The second
subject is the matching of a full and an effective field theory at low
energies in order to evaluate the effective Lagrangian in gauge
theories.  We compare the results of the independent low-energy
constants for the Standard Model with a heavy Higgs boson, obtained by
matching gauge-invariant Green's functions~\cite{EWChPT_gaugeinv},
with earlier evaluations that used different
methods~\cite{EWChPT_early, H_M_E_M_D_GK} and with those for a simple
Technicolor model~\cite{Technicolor}. We end with some comments on the
phenomenological implications of our investigation for strongly
interacting models of electroweak symmetry breaking.

%---------------------------------------------------------------------------

\vspace*{-2mm}
\section{The electroweak chiral Lagrangian}

Assuming a strongly interacting dynamics behind the breaking of the
electroweak symmetry, the physics at low energies can be described by
an effective Lagrangian, the electroweak chiral
Lagrangian~\cite{EWChPT} built in analogy to the chiral Lagrangian for
low-energy QCD~\cite{Weinberg79, GL84_85}. It is given by a series in
powers of momenta and masses, $\lag_{eff} = \lag_2 + \lag_4 + \ldots$,
where $\lag_k$ is of order $p^k$ and has the general form $\lag_k =
\sum_i \low_i^{(k)} {\cal O}_i^{(k)}$. The coefficients $\low_i^{(k)}$
represent the low-energy constants of the effective theory and count
as order $p^0$. The operators ${\cal O}_i^{(k)}$ involve the light
fields in such a way that they respect the $SU(2)_L \times U(1)_Y$
gauge symmetry. We assume that $p^2, M_W^2, M_Z^2 \ll M^2$, where $p$
is a typical momentum and $M$ is the mass scale of the heavy particles
in the underlying theory.

We first consider only the bosonic sector, i.e.\ the low-energy
dynamics of the $W$- and $Z$-boson and the photon. At lowest order,
the electroweak chiral Lagrangian~\cite{EWChPT} can be written in the
following form~\cite{EWChPT_gaugeinv}
\bea
\label{lag_2}
\lag_2 & = & {\bv^2\over 2} \left( \bW_\mu^+ \bW_\mu^- 
+ \trho {1\over 4} \bZ_\mu \bZ_\mu \right) 
+ {1\over 4\bgs}  \bW_{\mu\nu}^a \bW_{\mu\nu}^a 
+ {1\over 4\bgps} \bB_{\mu\nu} \bB_{\mu\nu}  \nonumber \\
& & -{1\over2} \bK_{\mu\nu} \bB_{\mu\nu} 
+ 2 \bv^2 (\bJ_\mu^+ \phase^+ \bW_\mu^- + \bJ_\mu^- \phase^-
\bW_\mu^+) + \bv^2 \bJ_\mu^\Z \bZ_\mu 
+ 4 c_\W \bv^2 \bJ_\mu^+ \bJ_\mu^-  
+ c_\Z \bv^2 \bJ_\mu^\Z \bJ_\mu^\Z  \ , 
\eea
with 
\bea 
\W_\mu^+ & = & i \, \tU^\dagger D_\mu U \ , \ \W_\mu^- = i \, U^\dagger
D_\mu \tU \ , \ \Z_\mu = - 2 i \, U^\dagger D_\mu U \ , 
\label{defWZ_gaugeinv} \\ 
\W_{\mu\nu}^a & = & \p_\mu \W_\nu^a - \p_\nu \W_\mu^a +
\varepsilon^{abc} \W_\mu^b \W_\nu^c  \ , \ B_{\mu\nu} = \p_\mu B_\nu - 
\p_\nu B_\mu \ , \\ 
D_\mu U & = & \left( \p_\mu - i {\tau^a \over 2} W_\mu^a - i {1\over 2}
B_\mu \right) U \label{covderiv} \ . 
\eea
The $SU(2)_L$ doublet $U$, confined to the sphere $U^\dagger U =1$,
describes the degrees of freedom of the Goldstone bosons. The
$Y$-charge conjugate doublet is defined by $\tU = i \tau_2 U^*$.
Since we will employ a functional approach we have coupled external
sources $K_{\mu\nu}, J_\mu^\pm$ and $J_\mu^\Z$ to gauge-invariant
fields in Eq.~(\ref{lag_2}). The phase factor $\phase^\pm$, which
appears in the source terms for the charged fields $\W_\mu^\pm$ in
Eq.~(\ref{lag_2}), is defined by
\be
\phase^\pm(x) = \exp\left({\mp i \int d^dy  \, \G_0(x-y) \,
\partial_\rho \B_\rho}(y) \right) \quad , \quad 
\G_0(x-y) = \langle x | {1 \over - \Box} | y \rangle \ , 
\ee
and ensures gauge invariance under the full group $SU(2)_L
\times U(1)_Y$. We will use Euclidean space notation throughout.  
%
%\noindent
At tree-level we have the relations  
\be \label{treelevelrelations} 
M_W^2 = {\bv^2 \bes \over 4 \bs^2} \ , \
M_Z^2 = \trho {\bv^2 \bes \over 4 \bs^2 \bc^2} \ , \
\bc^2 \equiv \cos^2 \theta_W = {\bgs \over \bgs + \bgps} \ , \ 
s^2 = 1 - c^2 \ , \ 
\bes = {\bgs \bgps \over \bgs + \bgps} \ , 
\ee
where $\trho = c^2 M_Z^2 / M_W^2$ is the inverse of the usual
$\rho$-parameter. 

In order to have a well-defined effective field theory a consistent
chiral momentum counting is required. Following the counting rules of
chiral perturbation theory~\cite{Weinberg79, GL84_85}, we treat the
Goldstone boson field $U$ and the phase factor $\phase^\pm$ as
quantities of order $p^0$. The momenta, the covariant derivative
$D_\mu$, the gauge fields $\W_\mu^\pm, \Z_\mu, B_\mu$, and the masses
$M_W$ and $M_Z$ are counted as order $p$. The consistency of these
rules requires that the coupling constants $g,g^\prime$ and $e$ count
as order $p$.  This is similar to the counting of $e$ as order $p$, if
virtual photons are included in chiral perturbation
theory~\cite{Urech}. The above rules imply that $c$ and $s$ are of
order $p^0$. Furthermore, we count $\trho-1$ in Eq.~(\ref{lag_2}) as
order $p^0$ in the low-energy expansion. Since $\trho - 1$ is
experimentally known to be very small~\cite{PDG_98}, the custodial
symmetry breaking term $(\trho-1) \Z_\mu Z_\mu$ is
sometimes~\cite{H_M_E_M_D_GK} only included at order $p^4$ in the
effective Lagrangian.  Finally, we count the external source
$K_{\mu\nu}$ as order $p^0$ and $J_\mu^\pm, J_\mu^\Z$ as order $p$.

The most general effective Lagrangian at order $p^4$ is given by
$\lag_4 = \lag_4^0 + \lag_4^s$, where the CP-even terms without
external sources can be written in the form $\lag_4^0 =
\sum_{i=1}^{18} \ourlow_i \ourop_i$. The operators $\ourop_i$ are
given by
\be \label{basis_lag_4}
\begin{array}{rclcrcl}
\ourop_{1} & = & (\bW_\mu^+ \bW_\mu^-) (\bW_\nu^+ \bW_\nu^-) \ , 
& \qquad & 
\ourop_{10} & = & i \bZ_\nu ( \bd_\mu \bW_\mu^+ \bW_\nu^- - \bd_\mu
\bW_\mu^- \bW_\nu^+ ) \ , \\ 
\ourop_{2} & = & (\bW_\mu^+ \bW_\nu^-) (\bW_\mu^+ \bW_\nu^-) \ , 
& \qquad & 
\ourop_{11} & = & \bZ_{\mu\nu} \bZ_{\mu\nu} \ , \\
\ourop_{3} & = & (\bZ_\mu \bZ_\mu) (\bW_\nu^+ \bW_\nu^-) \ , 
& \qquad & 
\ourop_{12} & = & \bB_{\mu\nu} \bZ_{\mu\nu} \ , \\
\ourop_{4} & = & (\bZ_\mu \bZ_\nu) (\bW_\mu^+ \bW_\nu^-) \ , 
& \qquad & 
\ourop_{13} & = & (\bd_\mu \bW_\mu^+) (\bd_\nu \bW_\nu^-) \ , \\
\ourop_{5} & = & (\bZ_\mu \bZ_\mu)(\bZ_\nu \bZ_\nu) \ , 
& \qquad & 
\ourop_{14} & = & (\p_\mu \bZ_\mu) (\p_\nu \bZ_\nu) \ , \\
\ourop_{6} & = & \epsilon_{\mu\nu\rho\sigma} \bZ_\sigma (\bW_\rho^-
\bW_{\mu\nu}^+ + \bW_\rho^+ \bW_{\mu\nu}^- ) \ , 
& \qquad & 
\ourop_{15} & = & M_W^2 \left( \bW_\mu^+ \bW_\mu^- +
{1\over4} \bZ_\mu \bZ_\mu \right) \ , \\ 
\ourop_{7} & = & i \bZ_{\mu\nu} (\bW_\mu^+ \bW_\nu^- - \bW_\nu^+
\bW_\mu^-) \ , 
& \qquad & 
\ourop_{16} & = & M_Z^2 \bZ_\mu \bZ_\mu \ , \\ 
\ourop_{8} & = & i \bB_{\mu\nu} (\bW_\mu^+ \bW_\nu^- - \bW_\nu^+
\bW_\mu^-) \ , 
& \qquad & 
\ourop_{17} & = & \bW_{\mu\nu}^a \bW_{\mu\nu}^a \ , \\
\ourop_{9} & = & i \bZ_\mu ( \bd_\mu \bW_\nu^+ \bW_\nu^- - \bd_\mu
\bW_\nu^- \bW_\nu^+) \ ,  
& \qquad & 
\ourop_{18} & = & \bB_{\mu\nu} \bB_{\mu\nu} \ , 
\end{array}
\ee
where
\be
\bd_\mu \W_\nu^\pm = (\partial_\mu \mp i B_\mu) \W_\nu^\pm \ , \ 
\W_{\mu\nu}^\pm = d_\mu \W_\nu^\pm - d_\nu \W_\mu^\pm \ , \
\Z_{\mu\nu} = \p_\mu \Z_\nu - \p_\nu \Z_\mu \ . 
\ee
One can easily identify those operators in Eq.~(\ref{basis_lag_4})
that contribute to the self-energies of the gauge bosons and to
anomalous triple and quartic gauge couplings, respectively.  We have
employed partial integrations to remove other allowed terms, like
$\bd_\mu \W_\nu^+ \bd_\nu \W_\mu^-$.  As shown in
Ref.~\cite{EWChPT_gaugeinv} the operators $\ourop_{1}, \ldots,
\ourop_{14}$ can be expressed through the operators $\op_1, \ldots,
\op_{14}$ in the usually employed basis~\cite{EWChPT}. The operator
$\ourop_{16}$ corresponds to the operator $\op_0$. Due to the factor
$M_Z^2$ this terms counts as order $p^4$. The CP-even source terms can
be written in the form $\lag_4^s = \sum_{i=1}^{76} l_i^s
\ourop_i^s$. The list of operators $\ourop_i^s$ can be found in
Ref.~\cite{EWChPT_gaugeinv}.

%---------------------------------------------------------------------------

\section{Number of independent parameters in $\lag_{eff}$}
\label{sec:independent_parameters}

The generating functional for the Green's functions of the
gauge-invariant fields $\phase^\mp \W_\mu^\pm, \Z_\mu$ and
$\B_{\mu\nu}$ is given by the path integral 
\be \label{gen_func}
  e^{-W_{eff}[\bK_{\mu\nu}, \bJ_\mu^\pm, \bJ_\mu^\Z}] = \int
\d\mu[\bU, W_\mu^a,\bB_\mu] \ e^{- \int  d^d x \, \lag_{eff}} \ . 
\ee
As discussed in Ref.~\cite{EWChPT_gaugeinv} it is possible to evaluate
the path integral in Eq.~(\ref{gen_func}) without fixing a gauge
because we consider only Green's functions of gauge-invariant fields.
In general, there are two different kinds of contributions to the
generating functional in Eq.~(\ref{gen_func}). On the one hand, one
has tree-level contributions given by the integral~$\int
d^dx\lag_{eff}$, which has to be evaluated at the stationary point,
i.e., with the solutions of the equations of motion. On the other hand
there are contributions from loops, which ensure unitarity. General
power counting arguments show that $n$-loop corrections are suppressed
by~${2n}$ powers of the momentum~\cite{Weinberg79, GL84_85}. For
instance, at order $p^2$ there are only tree-level contributions from
$\lag_2$, whereas at order $p^4$ one has to consider tree-graphs from
$\lag_4$ and one-loop graphs with vertices from $\lag_2$.

At order $p^2$, the generating functional of the effective field
theory is given by the classical action 
\be
W_2[\bK_{\mu\nu}, \bJ_\mu^\pm, \bJ_\mu^\Z] = \int d^dx \ \lag_2 
  \left(\bW_{\mu}^\pm, \Z_\mu, \bB_{\mu}; \bK_{\mu\nu}, \bJ_\mu^\pm,
\bJ_\mu^\Z \right) \ ,
\ee
where the gauge fields satisfy the equations of motion
\bea
- \bd_\mu \bW^\pm_{\mu\nu} 
 \!\!& = &\!\!- M_W^2 ( \W_\nu^\pm + 4 j^\pm_\nu )
\pm i ( \bZ_{\mu\nu} + \bB_{\mu\nu}) \bW^\pm_\mu \mp i
\bW^\pm_{\mu\nu} \bZ_\mu \mp i (\p_\mu \bZ_\mu) \bW^\pm_\nu \pm i
(\p_\mu \bZ_\nu) \bW^\pm_\mu \nonumber \\ % \label{eomWpm_eff} \\
& & \pm i \bZ_\nu \bd_\mu \bW^\pm_\mu \mp i
\bZ_\mu \bd_\mu \bW^\pm_\nu - (\bZ_\mu \bZ_\mu) \bW^\pm_\nu + (\bZ_\mu
\bZ_\nu) \bW^\pm_\mu \pm 2 \bW^\pm_\mu (\bW_\mu^+ \bW_\nu^- -
\bW_\nu^+ \bW_\mu^-)  \nonumber \\[-0.1cm] 
& & \label{eomWpm_eff} \\
- \p_\mu (\bZ_{\mu\nu} + \bB_{\mu\nu})\!\!& = &\!\!- \bc^2 M_Z^2
  ( \Z_\nu + {4 \over \trho} J^\Z_\nu ) - 4
  \Z_\nu \W_\mu^+ \W_\mu^- 
+ 2 \Z_\mu ( \W_\mu^+ \W_\nu^- + \W_\nu^+ \W_\mu^- )
\label{equmow3_eff} \\ 
& & + 2 i ( \W^+_{\mu\nu} \W^-_\mu  -
\W^-_{\mu\nu} \W^+_\mu ) - 2 i ( d_\mu\W_\mu^+\W_\nu^-  - 
d_\mu\W_\mu^-\W_\nu^+  
- d_\mu\W_\nu^+\W_\mu^- + d_\mu\W_\nu^-\W_\mu^+)  
\nonumber \\
- \p_\mu \bB_{\mu\nu}\!\!& = &\!\!\bs^2 M_Z^2 \PT_{\nu\mu} ( \Z_\mu +
{4 \over \trho} J^\Z_\mu )
- {\bes\over \bc^2} \p_\mu  \bK_{\mu\nu} \ .  \label{eomB_eff_2} 
\eea
The equations of motion (constraints) for the Goldstone bosons are
given by  
\bea
\bd_\mu \W^\pm_\mu &=& - 4 \bd_\mu j^\pm_\mu 
\pm i \bZ_\mu (\W_\mu^\pm + 4 j^\pm_\mu)
\mp i \trho (\Z_\mu + {4 \over \trho}
J^\Z_\mu) \bW^\pm_\mu \ , \label{eomU_pm_eff} \\ 
\p_\mu \Z_\mu &=& - {4\over \trho} \p_\mu
J^\Z_\mu + {8 i \over \trho} (\bW_\mu^+ 
\bj_\mu^- - \bW_\mu^- \bj^+_\mu )  \ , \label{eomU_Z_eff}
\eea
where $j_\mu^\pm = \phase^\pm J_\mu^\pm$ and $\PT_{\mu\nu} =
\delta_{\mu\nu} - {\p_\mu \p_\nu \over \Box}$.  As discussed in
Refs.~\cite{SM_gaugeinv, EWChPT_gaugeinv} it is possible to solve the
equations of motion~(\ref{eomWpm_eff}) -- (\ref{eomU_Z_eff}) for the
physical degrees of freedom without the need to fix a gauge.

The Lagrangian $\lag_4$ only contributes at tree-level if we go up to
order $p^4$ in the low-energy expansion of the generating functional
in Eq.~(\ref{gen_func}). Therefore we can use the equations of motion
for the gauge fields as well as the constraints for the Goldstone
bosons to eliminate redundant terms in the list of operators in
Eq.~(\ref{basis_lag_4}). Note that in our gauge-invariant approach no
gauge artefacts can enter through this procedure. Using the
constraints (\ref{eomU_pm_eff}) and (\ref{eomU_Z_eff}) one can remove
the following three operators
\bea
\ourop_{10} & = & - 2 (1-\trho) \ourop_4 + 4 \ourop_{4}^s - 4
\ourop_{6}^s - 4 \ourop_{46}^s  \ , \label{remove_O9} \\
\ourop_{13} & = & (1-\trho)^2 \ourop_4 - 4 (1-\trho) \ourop_{4}^s + 4
(1-\trho) \ourop_{6}^s + 16 \ourop_{14}^s - 16 \ourop_{17}^s
\nonumber \\
&&\mbox{} + 16 \ourop_{19}^s + 4 (1-\trho) \ourop_{46}^s - 16
\ourop_{51}^s + 16 \ourop_{53}^s + 16 \ourop_{74}^s \ ,
\label{remove_O13} \\ 
\ourop_{14} & = & (64 / \trho^2) (2 \ourop_{10}^s - \ourop_{12}^s)
+ (64 / \trho^2) (\ourop_{49}^s - \ourop_{52}^s) + (16 / \trho^2)
\ourop_{76}^s \ . \label{remove_O14} 
\eea
This corresponds to removing the operators $\op_{11},\op_{12},
\op_{13}$ in the usual basis, as was done in Ref.~\cite{EWChPT}. Using
the equations of motions for the gauge fields, Eqs.~(\ref{eomWpm_eff})
-- (\ref{eomB_eff_2}), one can remove the two operators
\bea
\ourop_{11} & = & - 8 \ourop_1 + 8 \ourop_2 - 16 \ourop_3 + 16 \trho
\ourop_4 + 8 \ourop_7 - 8 \ourop_9 - 8 \ourop_{15} 
- 2 \bc^2 (2 - 1 / \trho) \ourop_{16} - \ourop_{17} + \ourop_{18}
\nonumber \\
&&+ 32 \ourop_{4}^s - 32 \ourop_{6}^s - 32 \ourop_{46}^s - 16
\ourop_{64}^s - 16 (\bc^2 / \trho) \ourop_{66}^s  \ ,
\label{remove_O11} \\ 
\ourop_{12} & = & 8 \ourop_1 - 8 \ourop_2 + 8 \ourop_3 - 8 \trho
\ourop_4 - 4 \ourop_7 + 4 \ourop_9 + 8 \ourop_{15} 
+ 2 \bc^2 (1 - 1 / \trho) \ourop_{16} + \ourop_{17}
- \ourop_{18} \nonumber \\
&&- 16 \ourop_{4}^s + 16 \ourop_{6}^s + 16 \ourop_{46}^s + 16
\ourop_{64}^s + 8 (\bc^2 / \trho) \ourop_{66}^s \ . \label{remove_O12} 
\eea
This step, which corresponds to removing $\op_{1}$ and $\op_{8}$ in
the usual basis, has not been taken before in the literature. The
operators $\ourop_{11}$ and $\ourop_{12}$ contribute to the self-energies
of the gauge bosons which are not directly observable.  The relations
between the operators $\op_i$ in the usual basis which follow from
Eqs.~(\ref{remove_O9}) -- (\ref{remove_O12}) can be found in
Ref.~\cite{EWChPT_gaugeinv}. A similar procedure can be applied to
eliminate four of the source terms~\cite{EWChPT_gaugeinv}.

Furthermore, the operators $\ourop_{15}, \ldots, \ourop_{18}$ in
Eq.~(\ref{basis_lag_4}), as well as several source terms in
$\lag_4^s$, are proportional to corresponding terms in $\lag_2$. One
can remove these terms from the basis by a renormalization of the
parameters, low-energy constants and sources in $\lag_2$. For instance
we get
\be \label{v_eff}
{v^2 \over 2 } \left(\W_\mu^+ \W_\mu^- + {1\over 4} \Z_\mu \Z_\mu
\right) 
+ \ourlow_{15} M_W^2 \left(\W_\mu^+ \W_\mu^- + {1\over 4} \Z_\mu
\Z_\mu \right)  
\ \ \to \ \ {v_{eff}^2 \over 2} \left(\W_\mu^+ \W_\mu^- + {1\over 4}
\Z_\mu \Z_\mu \right)  \ . 
\ee
The complete list of redefinitions can be found in
Ref.~\cite{EWChPT_gaugeinv}.  

At the end we find~\cite{EWChPT_gaugeinv} that there are 9
independent low-energy constants $\ourlow_1^\prime,
\ldots, \ourlow_9^\prime$ in $\lag_4^0$.  Furthermore, there
are 63 independent low-energy constants $\ourlow_i^{s\prime}$ which
appear in the source terms $\lag_4^s$. Since the use of the
relations~(\ref{remove_O9}) -- (\ref{remove_O12}) modifies the
low-energy constants of the remaining terms, we have denoted the
independent constants by $\ourlow_i^\prime$ and
$\ourlow_i^{s\prime}$. We can cover the conventions used in
Ref.~\cite{H_M_E_M_D_GK}, where a custodial symmetry breaking term
proportional to $Z_\mu Z_\mu$ is included only in $\lag_4$, by setting
$\trho = 1$ and including the operator $\ourop_0 \doteq M_Z^2
\Z_\mu Z_\mu \equiv \ourop_{16}$ into the basis with the low-energy
constant $\ourlow_0^\prime$. In a purely bosonic effective field
theory the source terms will not contribute to physical quantities
(on-shell gauge bosons)~\cite{EWChPT_gaugeinv}. Therefore, apart from
the universal parameters $v^2, g$ and $g^\prime$ in $\lag_2$, there
are only 10 physically relevant low-energy constants up to order
$p^4$, namely $l_1^\prime, \ldots, l_9^\prime$ and $\trho -1$, or
$\ourlow_0^\prime$, depending on the momentum counting, to distinguish
different underlying theories.

%---------------------------------------------------------------------------

\section{Inclusion of fermions and the oblique parameters $S,T,U$}

In Ref.~\cite{EWChPT} the equations of motion for the gauge fields,
Eqs.~(\ref{eomWpm_eff}) -- (\ref{eomB_eff_2}), have not been employed
to find relations between operators in $\lag_4^0$. Therefore the set
of 12 low-energy constants in $\lag_4$ found in these references is
redundant.  Before any conclusions can be drawn, however, one has to
study the inclusion of fermions into the effective field
theory~\cite{EWChPT_gaugeinv}.  At order $p^2$ the fermionic part of
the effective Lagrangian contains several terms
\be \label{lag_2_f} 
\lag_2^f = \lag_2^{f, kin} + \lag_2^{f, Yuk} + \lag_2^{f, CC} +
        \lag_2^{f, NC} + \lag_2^{f, 4F} + \lag_2^{f, s} \ . 
\ee
They denote the kinetic terms, the Yukawa couplings, the coupling to
charged and neutral currents, four-fermion interactions and source
terms, respectively. In the following we will concentrate on the
charged current interactions, a more complete discussion can be found
in Ref.~\cite{EWChPT_gaugeinv}. These interactions can be expressed as
follows
\bea
 \lag_2^{f, CC} & = & 
        \sum_{ij} c_{CC}^{ij,L} \left( \bW_\mu^+ j_\mu^{L, ij
          -} + \bW_\mu^- j_\mu^{L, ij +} \right) 
        + c_{CC}^{ij,R} \left( \bW_\mu^+ j_\mu^{R, ij -} + 
 \bW_\mu^- j_\mu^{R, ij +} \right) \ , \label{lag_CC} \\ 
 j_\mu^{L/R, ij +} & = & \bar \bdf_{L/R}^i \gamma_\mu \bu_{L/R}^j
\quad , \quad  
 j_\mu^{L/R, ij -} =  \bar \bu_{L/R}^i \gamma_\mu \bdf_{L/R}^j  \ , 
\label{fermion_currents} 
\eea
where the fermionic currents are written in terms of $SU(2)_L$
gauge-invariant fields 
\be
        u_L^k = \tilde U^\dagger \Psi_L^k \quad , \quad 
        d_L^k = U^\dagger \Psi_L^k \ . 
\ee
Here, $\Psi^k$ denotes the usual fermion doublet for flavor number
$k$. The fields $u_R^k$ and $d_R^k$ in Eq.~(\ref{fermion_currents})
are the usual $SU(2)_L$ singlets. As can be seen from
Eq.~(\ref{lag_CC}), the effective Lagrangian contains a host of a
priori undetermined low-energy constants $c^{ij}$ and more appear in
$\lag_2^f$ in Eq.~(\ref{lag_2_f}) and at higher orders in $\lag_3^f,
\lag_4^f, \ldots $. Of course, there exist strong bounds on couplings
which deviate from the corresponding values in the Standard Model. In
a general effective Lagrangian analysis one has to include, however,
such terms and, in fact, they are expected to be small if they are of
higher order in the low-energy expansion.

Using the equations of motion for the Goldstone bosons (constraints)
one can still remove the terms $\ourop_{10}, \ourop_{13}$ and
$\ourop_{14}$. For instance, in the relation~(\ref{remove_O13}) we get
a contribution $(d_\mu j_\mu^+) (d_\nu j_\nu^-)$ from the source terms
which now contain the fermionic currents. Employing the equations of
motion for the fermions one observes that $(d_\mu j_\mu^\pm)$ is
proportional to the fermion masses which are very small for light
external fermions, cf.\ Ref.~\cite{EWChPT}.  On the other hand, if we
employ the equations of motion for the gauge fields we obtain, cf.\
Eq.~(\ref{remove_O11}),
\be
\ourop_{11} = \ldots - 16 M_W^2 (\W_\mu^+ j_\mu^- + \W_\mu^- j_\mu^+)
\ . 
\ee
The last term is proportional to a term in $\lag_2^{f, CC}$ and
therefore leads only to a non-observable renormalization of the
low-energy couplings $c^{ij}$ in Eq.~(\ref{lag_CC}), similarly to the
redefinition in Eq.~(\ref{v_eff}). Therefore, the operator
$\ourop_{11}$ can again be removed in the framework of a general
effective field theory approach. The same statement applies to
$\ourop_{12}$.

In contrast, in Ref.~\cite{EWChPT} the assumption was made that all
couplings of the fermions to the gauge bosons are identical to their
values in the Standard Model. In this case the use of the equations of
motion for the gauge fields does not reduce the number of parameters,
but simply shifts some low-energy constants from the bosonic to the
fermionic sector. The assumptions made in Ref.~\cite{EWChPT} might,
however, be too stringent. 

We have {\it chosen} to remove the operators $\ourop_{11}$ and
$\ourop_{12}$ from the basis, which correspond to the operators
$\op_1$ and $\op_8$ in the usual basis. Since these operators
contribute to the self-energies of the gauge bosons, the low-energy
constants $a_1$ and $a_8$ are sometimes identified with the oblique
correction parameters $S$ and $U$~\cite{STU}\footnote{The parameter
$T$ is often identified with the low-energy constant $a_0$ which
corresponds to $\trho-1$ or $\ourlow_0^{\prime}$, depending on the
momentum counting.}. As discussed in Ref.~\cite{EWChPT_gaugeinv}, we
believe that such an identification is not really possible. The
parameters $S,T$ and $U$, as used nowadays by the Particle Data
Group~\cite{PDG_98}, describe the effects of heavy new physics {\it
beyond} the Standard Model on the self-energies of the gauge
bosons. In particular, they depend on the Higgs boson mass $M_H$,
whereas the effective Lagrangian approach also works if there exists
no Higgs boson at all. In order to make contact between the two
descriptions one might do the following. 

The low-energy constants at order $p^k$ have the general form
$l_i^{(k)}= \delta_i^{(k)} \pole + l_i^{(k), r}(\mu)$. The
coefficients $\delta_i^{(k)}$ of the pole term, $\pole \doteq
{\mu^{d-4} \over 16\pi^2} \left( {1 \over d-4} - {1\over2} (\ln 4\pi +
\Gamma'(1) + 1) \right)$, are universal, i.e.\ independent of the
underlying theory. Following the conventions of chiral perturbation
theory~\cite{Weinberg79, GL84_85}, we denote the finite, scale
dependent parts, $l_i^{(k), r}(\mu)$, as renormalized low-energy
constants. One could now try to mimic any strongly interacting
symmetry breaking sector by studying the large Higgs mass limit. One
cannot completely remove the Higgs particle from the theory in this
way, however, due to non-decoupling effects for $M_H \to
\infty$. We now assume that the renormalized low-energy constants can
be decomposed as follows
\be \label{decompose}
  l_i^{(k), r}(\mu) = l_i^{(k), SM}(\mu) + l_i^{(k), new}(\mu) \ , 
\ee
where the first term describes the contributions for the Standard
Model with a heavy Higgs boson (see the next Section). The definition
of $S,T$ and $U$ now amounts to setting $l_i^{(k), new}(\mu) = 0$ for
all $i, k$ except for $k=4$ and $i=0, 11, 12$. This introduces three
finite parameters, independent of each other, to describe new physics
effects.  Again, the use of the equations of motion for the gauge
fields does not reduce the number of parameters in this case, but
simply shifts some low-energy constants from the bosonic to the
fermionic sector. The assumptions made in Ref.~\cite{STU}, i.e.\ that
new physics effects predominantly show up in oblique corrections and
that a decomposition according to Eq.~(\ref{decompose}) is possible in
the absence of a Higgs boson, might, however, again be too
restrictive.

%---------------------------------------------------------------------------

\section{Low-energy constants for a heavy Higgs boson and for a
simple Technicolor model}

The low-energy constants in the effective Lagrangian are determined by
the underlying theory of electroweak symmetry breaking and can, in
principle, be calculated by matching the full and the effective theory
at low energies. In the case of a heavy Higgs boson, which we consider
here as one illustrative example of a strongly interacting symmetry
breaking sector, the matching condition can be evaluated using
perturbative methods. One has to make sure, however, that no gauge
artefacts enter in this procedure, if one matches, for instance,
gauge-dependent Green's functions of gauge bosons~\cite{EWChPT_early,
H_M_E_M_D_GK}. In Ref.~\cite{EWChPT_gaugeinv} we have redone the
calculation within our manifestly gauge-invariant approach by matching
the generating functionals of our gauge-invariant Green's functions,
i.e.\ by requiring $W_{eff} = W_{SM}$. The matching condition was
evaluated up to one-loop order in the Standard Model (bosonic sector)
and up to order $p^4$ in the low-energy expansion.  At the level of
the bare effective Lagrangian we reproduce the results found in
Ref.~\cite{H_M_E_M_D_GK}. The complete result for the effective
Lagrangian $\lag_2$ after renormalization can be found in
Ref.~\cite{EWChPT_gaugeinv}. The independent low-energy constants of
the non-source terms in the bosonic sector at order $p^4$ are given by
\bea
\ourlow_{0}^\prime  & = & s_p^2 \left( {3\over 4} \pole + {3\over 8}
{1\over 16 \pi^2} \ln \left( {\MHps \over \mu^2} \right) + {1\over16} 
{1\over 16 \pi^2} \right) \ , \nonumber \\ 
\ourlow_1^\prime & = & - {1\over 3} \pole - {2 s_p^2 \MWps
    \over \ers \MHps} - {1\over 6} {1\over 16 \pi^2} \ln
\left( {\MHps \over \mu^2} \right) + {176 - 27
  \sqrt{3} \pi \over 36 \cdot 16 \pi^2} \ , \nonumber   \\ 
\ourlow_2^\prime & = & - {2\over 3} \pole - {1\over 3} {1\over
  16 \pi^2} \ln \left( {\MHps \over \mu^2} \right) - {1\over 18}
{1\over 16 \pi^2} \ , \nonumber \\
\ourlow_3^\prime & = & {1\over 6} \pole - {s_p^2 \MWps
    \over \ers \MHps} + {1\over 12} {1\over 16 \pi^2} \ln
\left( {\MHps \over \mu^2} \right) + {178 - 27
  \sqrt{3} \pi \over 72 \cdot 16 \pi^2} \ , \nonumber \\ 
\ourlow_4^\prime & = & - {2\over 3} \pole - {1\over 3} {1\over
  16 \pi^2} \ln \left( {\MHps \over \mu^2} \right) - {1\over 18}
{1\over 16 \pi^2} \ , \nonumber \\
\ourlow_5^\prime & = & - {1\over 16} \pole - {s_p^2 \MWps
    \over 8 \ers \MHps} - {1\over 32} {1\over 16 \pi^2} \ln
\left( {\MHps \over \mu^2} \right) + {58 - 9
  \sqrt{3} \pi \over 192 \cdot 16 \pi^2} \ , \nonumber \\
\ourlow_{6}^\prime & = & 0 \ , \nonumber \\ 
\ourlow_7^\prime & = & - {1\over 6} \pole - {1\over 12} {1\over
  16 \pi^2} \ln \left( {\MHps \over \mu^2} \right) - {13\over 72}
{1\over 16 \pi^2} \ , \nonumber \\
\ourlow_8^\prime & = & - {1\over 6} {1\over 16 \pi^2} \ , 
\nonumber \\
\ourlow_9^\prime & = & {1\over 6} \pole + {1\over 12} {1\over 16
  \pi^2} \ln \left( {\MHps \over \mu^2} \right) + {13\over 72} {1\over
  16 \pi^2} \ .
\label{l_i_SM} 
\eea
In the Standard Model the custodial symmetry breaking effects in
$\Delta \rho$ are proportional to $\bgps$. Therefore we have $\trho =
1$ in $\lag_2$ and there are 10 low-energy constants in $\lag_4$.  As
physical input parameters we have chosen the masses of the Higgs, and
the $W$- and $Z$-boson and the electric charge. The masses are defined
by the pole position of gauge-invariant two-point
functions~\cite{SM_gaugeinv} and are denoted by $M_{H,p}, M_{W,p}$ and
$M_{Z,p}$, respectively.  The electric charge, denoted by $e_{res}^2$,
is defined as the residue at the photon-pole of the two-point function
$\langle 0 | T B_{\mu\nu} B_{\rho\sigma} | 0 \rangle$. The value
agrees~\cite{SM_gaugeinv} with the usual result for the electric
charge in the Thompson limit.  Furthermore, we have introduced in
Eq.~(\ref{l_i_SM}) the notations $c_p^2 \doteq M_{W,p}^2 / M_{Z,p}^2,
\, s_p^2 = 1 - c_p^2$. The results for the low-energy constants are 
identical to those obtained in the ungauged $O(4)$-linear sigma model,
in all cases where such a comparison is possible. This can easily be
understood from our counting rules where $\gs, \gps =
\order(p^2)$. Gauge boson loops lead either to corrections of order 
$p^6$ or to terms in $\lag_4$ which are proportional to corresponding
terms in $\lag_2$ that are therefore not observable.  

For illustration, we have listed in Table~\ref{tab:l_i_SM} the values
of the renormalized low-energy constants $l_i^{\prime r}(\mu),
i=0,2,3,7$, for different scales $\mu$ and for several values of the
Higgs boson mass $M_{H,p}$. The low-energy constant $l_0^{\prime r}$
is related to $- \Delta \rho$. The low-energy constants $l_2^{\prime
r}$ and $l_3^{\prime r}$ represent anomalous quartic couplings of the
gauge bosons, whereas $l_7^{\prime r}$ contributes to anomalous triple
gauge couplings. Note that $l_4^{\prime r} = l_2^{\prime r}$ and
$l_9^{\prime r} = - l_7^{\prime r}$. Furthermore, $l_6^{\prime r}$
vanishes and $l_8^{\prime r}$ is independent of $M_{H,p}$ and
$\mu$. The low-energy constants $l_1^{\prime r}, l_3^{\prime r}$ and
$l_5^{\prime r}$ receive tree-level contributions which dominate
numerically for Higgs masses below $1~TeV$. Thus we have in this
region $l_1^{\prime r} \approx 2 l_3^{\prime r} \approx 16 l_5^{\prime
r}$. For Higgs masses above $1~TeV$ the application of perturbation
theory is certainly questionable.  Therefore, the values of the
low-energy constants quoted for $M_{H,p} = 2~TeV$ are only given as
illustration.  On the other hand, if the Higgs mass is too small,
e.g.\ below $250~GeV$, the low-energy expansion will break down.

\vspace*{-4mm}
\begin{table}[ht]
\caption{Values of some of the renormalized low-energy constants
$l_i^{\prime r}(\mu)$ for the Standard Model at different scales and
for several values of the mass of the heavy Higgs boson.}
\begin{center}
\renewcommand{\arraystretch}{1.2}
%\hspace*{-1.5mm}
\begin{tabular}{|c|c|c|c|c||c|c|c|c||c|c|c|c|}
\hline
 & \multicolumn{4}{|c||}{
 {$10^3 \times l_i^{\prime r}(\mu=M_Z)$}} & 
 \multicolumn{4}{|c||}{
 {$10^3 \times l_i^{\prime r}(\mu=1~TeV)$}} & 
 \multicolumn{4}{|c|}{
 {$10^3 \times l_i^{\prime r}(\mu=2~TeV)$}} 
\\ \hline
 $\!\!M_{H,p}[GeV]\!\!\!$ & 
250 & 500 & 1000 & 2000 & 
250 & 500 & 1000 & 2000 & 
250 & 500 & 1000 & 2000 
\\ \hline
$l_0^{\prime r} $ 	& 1.15	& 1.89 	& 2.62 	& 3.35	
			& -1.38	& -0.64 & $\!0.088\!$ & 0.82 	 
			& -2.11 & -1.38 & -0.64 & $\!0.088\!$ \\      
$l_2^{\prime r} $ 	& -4.61 & -7.54	& -10.5	& -13.4  
			& 5.50 	& 2.57	& -0.35	& -3.28 
			& 8.43	& 5.50 	& 2.57 	& -0.35 \\
$l_3^{\prime r} $   	& -247  & -58.2 & -10.4 & 2.07	
			& -250	& -60.7	& -12.9 & -0.46
			& -250	& -61.5	& -13.7	& -1.19 \\
$l_7^{\prime r} $ 	& -2.21	& -2.94	& -3.67	& -4.40	
			& 0.32	& -0.41	& -1.14	& -1.87
			& 1.05	& 0.32	& -0.41	& -1.14 \\
\hline
\end{tabular}   
\label{tab:l_i_SM}  
\end{center}
\end{table}

For comparison, in a two-flavor QCD-like Technicolor
model~\cite{Technicolor} one obtains the following estimates for the
independent low-energy constants\footnote{In the usually employed
basis one obtains the following results for the non-vanishing
renormalized low-energy constants at the scale $\mu = M_{\rho_{TC}}:
- (2 / 3) a_1^r = - a_2^r = a_3^r = 2 a_4^r = - 2 a_5^r = v^2 / (4
M^2_{\rho_{TC}})$. After eliminating the operators $\op_1$ and $\op_8$
one gets: $- a_2^{\prime r} = - 2 a_3^{\prime r} = 2 a_4^{\prime r} =
- 2 a_5^{\prime r} = v^2 / (4 M^2_{\rho_{TC}})$ at the scale $\mu =
M_{\rho_{TC}}$.}
\bea
\ourlow_1^{\prime r}(\mu = M_{\rho_{TC}}) & = & {3 \over 4} {v^2 \over
M^2_{\rho_{TC}}}  \ , \hspace*{0.4cm}  
\ourlow_2^{\prime r}(\mu = M_{\rho_{TC}}) \hspace*{0.12cm} = \; - {3
\over 4} {v^2 \over M^2_{\rho_{TC}}}  \ , \hspace*{0.3cm} 
\ourlow_3^{\prime r}(\mu = M_{\rho_{TC}}) \; = \; {3 \over 4} {v^2 \over
M^2_{\rho_{TC}}}  \ , \nonumber \\ 
\ourlow_4^{\prime r}(\mu = M_{\rho_{TC}}) & = & - {3 \over 4} {v^2 \over
M^2_{\rho_{TC}}}  \ , \ 
\ourlow_7^{\prime r}(\mu = M_{\rho_{TC}}) \, \, = \; - {1 \over 4}
{v^2 \over M^2_{\rho_{TC}}}  \ , \hspace*{0.3cm}  
\ourlow_8^{\prime r}(\mu = M_{\rho_{TC}}) \; = \; {1 \over 8} {v^2 \over
M^2_{\rho_{TC}}}  \ , \nonumber \\ 
\ourlow_9^{\prime r}(\mu = M_{\rho_{TC}}) & = & {1 \over 4} {v^2 \over
M^2_{\rho_{TC}}}  \ , \hspace*{0.4cm}  
\ourlow_i^{\prime r}(\mu = M_{\rho_{TC}}) \; = \; 0, \quad i = 0, 5, 6, 
\label{l_i_TC} 
\eea
where $M_{\rho_{TC}}$ denotes the mass of the Technirho meson.  We
note that these expressions have not been obtained through a matching
calculation. Instead, they are taken over from the corresponding
estimates for the low-energy constants in the ordinary chiral
Lagrangian, obtained by assuming that the exchange of the lowest lying
resonances dominates the numerical values of the renormalized
low-energy constants in the resonance region, i.e.\ at the scale of
the Technirho mass. This assumption works reasonably well for the
coefficients in the chiral Lagrangian for QCD~\cite{GL84_85,
ChPT_Resonances} and can be justified using large-$N_c$ arguments and
constraints from sum-rules~\cite{largeNc_LMD}. By taking over these
estimates from QCD we assume that the electroweak interactions
(gauge-boson loops) will not drastically change the corresponding
values of the low-energy constants in the electroweak chiral
Lagrangian. In contrast to the matching calculation for the Standard
Model with a heavy Higgs boson this cannot be shown explicitly in the
Technicolor model, because of the strongly interacting dynamics.
In any case, we will use the values given in Eq.~(\ref{l_i_TC}) only
for illustration, since QCD-like Technicolor models are
phenomenologically ruled out due to large flavor-changing
neutral-currents. 

In Table~\ref{tab:l_i_SM_TC} we have collected the values of all
renormalized low-energy constants $l_i^{\prime r}(\mu)$ for the
Standard Model with a heavy Higgs boson, for $M_{H,p} = 250~GeV$ and
$1000~GeV$, and for the Technicolor model. We have chosen two
representative scales, the mass of the $Z$-boson and the mass of the
Technirho meson $M_{\rho_{TC}}$.  Simply scaling up from QCD we used 
$M_{\rho_{TC}} = 2~TeV$ and $v = 246~GeV$.

\vspace*{-4mm}
\begin{table}[ht]
\caption{Values of the renormalized low-energy constants $l_i^{\prime
r}(\mu)$ for the Standard Model and the two-flavor Technicolor (TC)
model at the scale of the $Z$-boson mass and at the mass-scale of the
Technirho. For the Standard Model case two values for the mass of the
heavy Higgs boson have been chosen.}
\begin{center}
\renewcommand{\arraystretch}{1.1}
\begin{tabular}{|c|r@{.}l|r@{.}l|r@{.}l||r@{.}l|r@{.}l|r@{.}l|}
\hline
 & \multicolumn{6}{|c||}{
 {$10^3 \times l_i^{\prime r}(\mu=M_Z)$}} & 
 \multicolumn{6}{|c|}{
 {$10^3 \times l_i^{\prime r}(\mu=M_{\rho_{TC}})$}} 
 \\ \cline{2-13} 
 & \multicolumn{2}{|c|}{{$\!M_{H,p} = 250~GeV\!$}} 
 & \multicolumn{2}{|c|}{{$\!M_{H,p} = 1000~GeV\!$}} 
 & \multicolumn{2}{|c||}{TC} 
 & \multicolumn{2}{|c|}{{$\!M_{H,p} = 250~GeV\!$}} 
 & \multicolumn{2}{|c|}{{$\!M_{H,p} = 1000~GeV\!$}} 
 & \multicolumn{2}{|c|}{TC} 
 \\ \hline
% Values for v=246
$l_0^\prime$ & \hspace*{1cm} 1 & 15 	& \hspace*{1cm} 2 & 62  
& 3 & 26  & \hspace*{1cm}-2 & 11 & \hspace*{1cm}-0 & 64 
&  0 & 0 \\       
$l_1^\prime$ & -498 & 8 & -31 & 3 & 4 & 83  & -492 & 3 & -24 & 8	
& 11 & 3 \\      
$l_2^\prime$ & -4 & 61 	& -10 & 5 & -24 & 4 & 8 & 43   & 2 & 57	
& -11 & 3 \\ 
$l_3^\prime$ & -247 & 1 & -10 & 4 & 14 & 6  & -250 & 4 & -13 & 7 
& 11 & 3 \\      
$l_4^\prime$ & -4 & 61 	& -10 & 5 & -24 & 4 & 8 & 43   & 2 & 57 	
& -11 & 3 \\
$l_5^\prime$ & -31 & 5 	& -2 & 61 & -1 & 22 & -30 & 2  & -1 & 39	
& 0 & 0 \\       
$l_6^\prime$ & 0 & 0  	& 0 & 0   & 0 & 0   & 0 & 0    & 0 & 0  	
& 0 & 0 \\       
$l_7^\prime$ & -2 & 21 	& -3 & 67 & -7 & 04 & 1 & 05   & -0 & 41	
& -3 & 78 \\
$l_8^\prime$ & -1 & 06 	& -1 & 06 & 1 & 89  & -1 & 06  & -1 & 06   
& 1 & 89 \\   
$l_9^\prime$ & 2 & 21 	& 3 & 67  & 7 & 04  & -1 & 05  & 0 & 41 	
& 3 & 78 \\ 
\hline
\end{tabular}
\label{tab:l_i_SM_TC}  
\end{center}
\end{table} 

\vspace*{-2mm}
One has to evolve the estimates for the low-energy constants for the
Technicolor model in Eq.~(\ref{l_i_TC}) from the resonance region down
to the scale of the $Z$-boson in order to compare with LEP~I / SLC
physics. In this way the constant $\ourlow_0^{\prime r}(\mu = M_Z)$
turns out to be nonzero, even though the Technicolor model preserves
the custodial symmetry. We observe that at both scales the pattern of
the low-energy constants for the Standard Model with a heavy Higgs
boson and for the Technicolor model is very different. Therefore, it
may be misleading to mimic any strongly interacting symmetry breaking
sector by a heavy Higgs boson, as is done, for instance, when fitting
values for the oblique parameters $S,T$ and $U$ in order to get bounds
on Technicolor models~\cite{PDG_98, Erler}.

We cannot directly compare the values for the low-energy constants
quoted in Table~\ref{tab:l_i_SM_TC} with experimental data, since we
consider here only the bosonic part of the effective
Lagrangian. Nevertheless, one might be tempted to make the
identification $\widehat \alpha(M_Z) T = - \widehat l_0^{\prime r}(\mu
= M_Z)$, cf.\ Eq.~(\ref{treelevelrelations}). Quantities with a hat
are defined in the $\overline{MS}$-scheme. A recent fit of the
parameters $S,T$ and $U$ yields $T = 0.00 \pm
0.15$~\cite{Erler}. Using the relation $l_0^{\prime r}(\mu) = \widehat
l_0^{\prime r}(\mu) + (3 s_p^2) / (8 \cdot 16 \pi^2)$, this would
translate to the bounds $- 1.8 < 10^3 \times l_0^{\prime r}(\mu = M_Z)
< 2.9$ at the $2\sigma$-level.  In Ref.~\cite{Erler} and in the
approach adopted by the Particle Data Group~\cite{PDG_98} the
parameter $T$ is equivalent to the parameter $\rho_0$, using $\widehat
\alpha(M_Z) T \equiv 1 - \rho_0^{-1}$, where $\rho_0 \equiv M_W^2 /
(M_Z^2 \widehat c_Z^2 \widehat \rho(m_t,M_H))$.  Fitting the
electroweak data with $\rho_0$ as a free parameter leads to
$\rule[-1mm]{0mm}{5mm} \rho_0 = 0.9996 {+ 0.0031 \atop - 0.0013}$ at
the $2\sigma$-level~\cite{Erler}. From this we would obtain the bounds
$- 2.2 < 10^3 \times l_0^{\prime r}(\mu = M_Z) < 2.2$.  Note, however,
that these fits for $\rho_0$ or $T$ are usually performed by including
radiative corrections within the context of the Standard
Model~\cite{PDG_98, Erler}. These quantities depend therefore in a
particular way on the masses of the top quark and of the Higgs boson.
These dependencies may change drastically, however, in other models
where $\rho \neq 1$ at tree level, if the full radiative corrections
of that model are taken into account, see
Ref.~\cite{Czakon_etal}. Furthermore, the quantities $\rho_0$ or $T$
are defined in Refs.~\cite{PDG_98, Erler} to parametrize only new
physics effects {\it beyond} the Standard Model. Therefore, a
reference value for the Higgs boson mass has to be chosen for the fit
of $S,T$ and $U$, e.g.\ for the value for $T$ quoted above, $M_H =
600~GeV$ has been used to mimic Technicolor models. In contrast, the
low-energy constant $l_0^{\prime r}(\mu = M_Z) \equiv - s_p^2
a_0^{\prime r}(\mu = M_Z)$ corresponds to the parameter $- \Delta
\rho$, see Ref.~\cite{H_M95} for the exact relation.
Within the Standard Model with a heavy Higgs boson we obtain from
$\ourlow_0^{\prime r}$ in Eq.~(\ref{l_i_SM}) the known Higgs
contribution to $\Delta \rho$. In other models, however, the
low-energy constant $\ourlow_0^{\prime r}$ may receive contributions
that are quite different from the Standard Model or from the
Technicolor model. Therefore we conclude that there exist no strong,
model-independent bounds on the low-energy constant $\ourlow_0^{\prime
r}$ and thus on models for a strongly interacting symmetry breaking
sector using current electroweak precision data. A similar conclusion
has been drawn recently in Ref.~\cite{BaggerFalkSwartz} where it was
argued that the electroweak precision data cannot rule out a strongly
interacting symmetry breaking sector with a scale of new physics as
high as $3~TeV$.

The strongest bounds for anomalous triple gauge couplings from
LEP~II~\cite{TGV_bounds} translate at the $2\sigma$-level to $10^3
\times | \ourlow_i^{\prime r} | \lesssim 100, i=7,8,9$. Therefore they
do not really constrain at the moment models for a strongly
interacting symmetry breaking sector where the size of the
renormalized low-energy constants is expected to be of the order of
${1\over 16 \pi^2}$~\cite{dim_analysis}.  All low-energy constants in
Table~\ref{tab:l_i_SM_TC} are of this order of magnitude, except for
$l_i^{\prime r}, i=1,3,5$, which receive large tree-level
contributions in the Standard Model.

%---------------------------------------------------------------------------

\section{Conclusions} 

We have reported here on our recent
investigation~\cite{EWChPT_gaugeinv} of the electroweak chiral
Lagrangian, which is based on the gauge-invariant approach to the
Standard Model introduced in Ref.~\cite{SM_gaugeinv}. For a purely
bosonic effective field theory we have shown, by using the equations
of motion and by redefining the universal parameters in the lowest
order Lagrangian $\lag_2$, that there are only 10 independent
low-energy constants in the effective Lagrangian up to order $p^4$.
In particular, it is possible to remove two operators that contribute
to the self-energies of the gauge bosons.  Hence, there is only one
low-energy constant in the effective Lagrangian, related to
$\Delta \rho$, that parametrizes oblique corrections. 
This result persists in the presence of fermions in a general
effective field theory analysis, where, however, many more new
low-energy constants arise in the fermionic sector. 
We briefly discussed the relation of this result to those found in the
usual approach to the electroweak chiral Lagrangian in
Ref.~\cite{EWChPT} where the couplings of the fermions are assumed to
have their Standard Model values. Furthermore we pointed out that the
oblique parameters $S,T$ and $U$~\cite{STU}, as determined by the fits
in Ref.~\cite{PDG_98, Erler}, cannot directly be identified with the
low-energy constants in the effective Lagrangian. They parametrize new
physics effects {\it beyond} the Standard Model and depend, in
particular, on the Higgs boson mass, whereas the effective Lagrangian
approach also works if there exists no Higgs boson at all.

We then discussed the evaluation of the effective Lagrangian by
matching gauge-invariant Green's function at low energies. For the
case of the Standard Model with a heavy Higgs boson the matching
relation can be evaluated in perturbation theory. With our approach we
reproduce the results of earlier calculations~\cite{EWChPT_early,
H_M_E_M_D_GK}. This can be understood from the momentum counting rule
$\gs, \gps = \order(p^2)$. We then compared the values of the
independent low-energy constants with those for a simple Technicolor
model. We find that the pattern of the low-energy constants is very
different in these two models. Therefore it may be misleading to
mimic any strongly interacting symmetry breaking sector with a heavy
Higgs boson, e.g.\ when fitting the oblique parameters $S,T$ and
$U$~\cite{PDG_98, Erler}. Note that there are non-decoupling effects,
if one tries to send the Higgs mass to infinity. Since, in addition,
all these fits are based on the inclusion of radiative corrections
within the context of the Standard Model, which may be quite
misleading, cf.\ Ref.~\cite{Czakon_etal}, we believe that the current
electroweak precision data do not really rule out models for a
strongly interacting symmetry breaking sector, see also
Ref.~\cite{BaggerFalkSwartz}.

If the electroweak symmetry breaking is indeed governed by a strongly
interacting underlying dynamics it will be one of the tasks to be
performed at a linear collider to determine these low-energy constants
in the effective Lagrangian experimentally as precisely as
possible~\cite{Han}, especially, if the collider is running at
energies below the resonance region.

%---------------------------------------------------------------------------

\vspace*{-2mm}
\section*{Acknowledgements}

I would like to thank A.~Schenk for the pleasant collaboration on the
topics presented here. Furthermore, I am indebted to J.~Gasser,
F.~Jegerlehner, M.~Knecht, H.~Leutwyler, E.\ de~Rafael,
V.~Ravindran. T.~Riemann, R.~Sommer, J.~Stern, and A.~Vicini for
useful discussions. This work was supported in part by Schweizerischer
Nationalfonds.

%---------------------------------------------------------------------------
% References 

%\newpage


\begin{thebibliography}{44}

\bibitem{SM_gaugeinv}
\preprintref
        {A.\ Nyf\/feler and A.\
          Schenk}{DESY-Preprint}{DESY-98-202}{December 1998;
          hep-ph/9812437}; to appear in Phys.\ Rev.\ D. 


\bibitem{EWChPT_gaugeinv}
\preprintref
        {A.\ Nyf\/feler and A.\
          Schenk}{DESY-Preprint}{DESY-99-088}{July 1999;
          hep-ph/9907294}.


\bibitem{EWChPT_early}
\artref
        {T.\ Appelquist and C.\ Bernard}{Phys.\ Rev.\ D}{22}{200}{1980};
%
\bookref
        {T.\ Appelquist}{{\rm in} Gauge Theories and Experiments at High
        Energies, {\rm edited by K.\ C.\ Bowler and D.\ G.\
          Sutherland}}{Scottish University Summer School in Physics,
        St.\ Andrews, 1980}.
%
\artref
        {A.\ C.\ Longhitano}{Phys.\ Rev.\ D}{22}{1166}{1980};
%
        Nucl.\ Phys.\ B {\bf 188}, 118 (1981).
%%


\bibitem{EWChPT}
\artref
        {T.\ Appelquist and G.-H.\ Wu}{Phys.\ Rev.\
          D}{48}{3235}{1993};
%
        ibid.\ {\bf 51}, 240 (1995); 
%
\artref
        {F.\ Feruglio}{Int.\ J.\ Mod.\ Phys.\ A}{8}{4937}{1993}. 


\bibitem{Han}
T.\ Han, Plenary talk presented at the 5th International Linear Collider
Workshop, Sitges, Spain, 28~April - 5~May 1999; hep-ph/9910495 and
references therein. 


\bibitem{H_M_E_M_D_GK}
\artref
        {M.\ J.\ Herrero and E.\ R.\ Morales}{Nucl.\
          Phys.\ B}{418}{431}{1994};
%
        ibid.\ {\bf 437}, 319 (1995); 
%
\artref
        {D.\ Espriu and J.\ Matias}{Phys.\ Lett.\ B}{341}{332}{1995}; 
%
\artref
        {S.\ Dittmaier and C.\ Grosse-Knetter}{Nucl.\
          Phys.\ B}{459}{497}{1996};  % \\
%        
\artref
        {J.\ Matias}{Nucl.\ Phys.\ B}{478}{90}{1996}. % \\


\bibitem{Technicolor}
\artref
        {S.\ Weinberg}{Phys.\ Rev.\ D}{19}{1277}{1979};
%
\artref
        {L.\ Susskind}{Phys.\ Rev.\ D}{20}{2619}{1979};
%
\artref
        {E.\ Farhi and L.\ Susskind}{Phys.\ Rep.}{74}{277}{1981}.


\bibitem{Weinberg79}
\artref
        {S.\ Weinberg}{Physica A}{96}{327}{1979}. 


\bibitem{GL84_85}
\artref
        {J.\ Gasser and H.\ Leutwyler}{Ann.\ Phys.}{158}{142}{1984};
%
        Nucl.\ Phys.\ B {\bf 250}, 465 (1985).


\bibitem{Urech}
\artref
      {R.\ Urech}{Nucl.\ Phys.\ B}{433}{234}{1995}.


\bibitem{PDG_98}
{\rm J.\ Erler and P.\ Langacker, in {\it Review of Particle
Physics}, Eur.\ Phys.\ J.\ C} {\bf 3}, {\rm 90 (1998)}.


\bibitem{Erler}
For a recent update see: J.\ Erler, Talk presented at the American
Physical Society (APS) Meeting of the Division of Particles and Fields
(DPF 99), Los Angeles, USA, 5 - 9 January 1999; hep-ph/9903449.


\bibitem{STU}
\artref
        {M.\ E.\ Peskin and T.\ Takeuchi}{Phys.\ Rev.\
          Lett.}{65}{964}{1990};
%
        Phys.\ Rev.\ D {\bf 46}, 381 (1992).


\bibitem{ChPT_Resonances}
\artref
	{G.\ Ecker, J.\ Gasser, A. Pich and E. de Rafael}{Nucl.\
Phys.\ B}{321}{311}{1989}; 
%
\artref
	{G.\ Ecker, J.\ Gasser, H.\ Leutwyler, A.\ Pich and E.\ de
Rafael}{Phys.\ Lett.\ B}{233}{425}{1989}.  


\bibitem{largeNc_LMD}
\artref
	{S.\ Peris, M.\ Perrottet and E.\ de
Rafael}{JHEP}{05}{011}{1998}; hep-ph/9805442.  


\bibitem{Czakon_etal}
\artref
	{M.\ Czakon, J.\ Gluza, F.\ Jegerlehner and M.\
Zra\l ek}{Eur.\ Phys.\ J.\ C}{13}{275}{2000}. 
%hep-ph/9909242 


\bibitem{BaggerFalkSwartz} 
\artref
	{J.A.\ Bagger, A.F.\ Falk and M.\ Swartz}{Phys.\ Rev.\
Lett.}{84}{1385}{2000}. 
%hep-ph/9908327. 


\bibitem{H_M95}
See the second paper in Ref.~\cite{H_M_E_M_D_GK}. 


\bibitem{TGV_bounds}
J.\ Mnich, Plenary talk given at the International Europhysics
Conference on High-Energy Physics: EPS-HEP 99, Tampere, Finland, 15 -
21 July 1999.
% 
%TESTS OF THE STANDARD MODEL.
%By J. Mnich (CERN). CERN-EP-99-143, Oct 1999. 31pp. 
%CERN Library Only. 


\bibitem{dim_analysis}
\artref
     {A.\ Manohar and H.\ Georgi}{Nucl.\ Phys.\ B}{234}{189}{1984}.

\end{thebibliography}
\end{document}